\documentclass[useAMS,usenatbib]{mn2e}
\usepackage{graphicx}
\usepackage{eufrak}
\usepackage{eucal}
\usepackage{txfonts}

\begin{document}

\title[ESO 546-G34: The most metal poor LSB galaxy?]{ESO 546-G34: The most metal poor LSB galaxy?}

\author[Mattsson, Pilyugin \& Bergvall]
       {Lars~Mattsson$^{1}$,
        Leonid~S.~Pilyugin$^{2}$
        and Nils~Bergvall$^{3}$\\
     $^{1}$ Dark Cosmology Centre, Niels Bohr Institute,
            University of Copenhagen, Juliane Maries Vej 30,
            DK-2100, Copenhagen \O, Denmark\\
     $^{2}$ Main Astronomical Observatory
            of the National Academy of Sciences of Ukraine,
            27 Zabolotnogo str., 03680 Kiev, Ukraine \\
     $^{3}$ Department of Physics and Astronomy, Uppsala University,
            Box 516,
            751 20 Uppsala, Sweden \\
             }

\date{}

\pagerange{\pageref{firstpage}--\pageref{lastpage}} \pubyear{2011}

\maketitle

\label{firstpage}

\begin{abstract}
We present a re-analysis of spectroscopic data for 23 H~\textsc{ii}-regions in 12 blue, metal-poor low surface brightness galaxies (LSBGs) taking advantage of recent 
developments in calibrating strong-line methods. In doing so we have identified a galaxy (ESO 546-G34) which may be the most metal-poor LSB galaxy found in the 
local Universe. Furthermore, we see evidence that blue metal-poor LSBGs, together with {blue compact galaxies (BCGs)} and many other H~\textsc{ii} galaxies, fall
outside the regular luminosity-metallicity relation. This suggests there might be an evolutionary connection between LSBGs and BCGs. In such case, several very 
metal-poor LSBGs should exist in the local Universe.
\end{abstract}


\section{Introduction}
Among blue metal-poor low surface brightness galaxies (LSBGs) one may very well find the most unevolved objects in the low-$z$ Universe \citep{Bothun90,Lacey91}.
LSBGs investigated by \citet{Ronnback94,Ronnback95} as well as those studied by \citet{deBlok96} tend to be bluer than most other galaxies of the same type. 
However, the wide range of colours and metallicities, from very blue and metal-poor to very red and relatively metal-rich, suggests that LSBGs also have a wide 
range of evolutionary states, just like their high surface brightness (HSB) counterparts \citep{Mattsson08,Liang10}. It is then natural 
to assume that blue LSB galaxies are young and unevolved and red LSB galaxies are old and evolved, especially since the blue ones appear metal poor and the
red ones quite metal rich \citep{McGaugh94, Kuzio04}.

A simple and attractive explanation for the LSB property is that LSBGs are simply forming stars very inefficiently \citep{Bothun90, Boissier08}. This inefficiency 
may reflect lower star-formation rates in the star-forming regions in general, or a star formation threshold leading to fewer star-forming regions, i.e., 
a lower star formation density in total \citep[see, e.g.,][]{Kennicutt89,vanderHulst93}. This picture is also very much consistent with the rather high gas mass fractions found 
in LSBGs \citep[see, e.g.][and references therein]{Kuzio04}. If this hypothesis for the LSB property holds true, a direct consequence would be that blue LSBGs are very
unevolved objects and some of them may have extremely low metallicities.

%Unevolved LSBGs have been suggested as the progenitors of blue compact galaxies (BCGs), where the latter then is formed out of a merger between a blue (dwarf)
%LSBG and another dwarf-type galaxy \citep{Bergvall99,Bergvall00}. There is, in fact, evidence that H~\textsc{ii} galaxies may have formed in such a way in general 
%\citep{Taylor96,Taylor97} .  If also extremely metal-poor BCGs, such as the famous I Zw 18, are the result of mergers involving a LSBGs, those LSBGs must have been 
%just as metal poor.

It is often difficult to detect the ${\rm [O~\textsc{iii}]} \lambda 4363$-line in LSBGs and hence impossible to use a 'direct' {method (based on the electron temperature 
$T_{\rm e}$)} to derive abundances. This suggests there might exist extremely metal-poor LSBGs which have not been labelled as such, since a $T_{\rm e}$-based 
method could not be used and existing strong-line calibrations are not made for the very metal-poor regime or at best give very uncertain results. 

In this letter we reanalyse, in the light of recent developments in calibrating strong-line methods of high-precision {\citep{Pilyugin11,Pilyugin10}}, a sample of blue 
unevolved LSBGs observed spectroscopically by \citet{Ronnback95}. Our results indicate that one of the expected extremely metal-poor LSBGs is present in our sample. 
{We also compare the rederived LSBG abundances to the abundances of metal-poor BCGs and H~\textsc{ii} galaxies}

  \begin{table}
  \begin{center}
  \caption{\label{indices} Strong-line indices and H$\beta$ fluxes for the 23 H~\textsc{ii}-regions in this study.}
  \begin{tabular}{lccccc}
  Object 
  & $R_2$ 
  & $R_3$
  & $N_2$
  & $S_2$
  & I(H$\beta)\cdot 10^{19}$ \\ 
  & & & & & [W m$^{-2}$]\\
  \hline
  %146G14A  &    6.52 &      2.99 &      -  &     2.070\\
  146G14B  &    2.23 &      3.52 &       0.146 &     0.430 & 6.59\\
  146G14C  &    3.26 &      3.56 &       0.200 &     0.620 & 21.0\\[1.5mm]
  489G56A  &    0.74 &      4.51 &       0.044 &     0.138 & 5.90\\
  489G56B  &    1.26 &      1.88 &       0.106 &     0.295 & 2.44\\[1.5mm]
  505G04A  &    3.06 &      2.90 &       0.153 &     0.523 & 9.13\\
  505G04B  &    2.89 &      3.06 &       0.170 &     0.498 & 8.89\\[1.5mm]
  546G34A  &    1.79 &      2.20 &       0.062 &     0.304 & 3.83\\
  546G34B  &    1.81 &      1.45 &       0.069 &     0.286 & 2.82\\[1.5mm]
  462G22   &    2.94 &      2.71 &       0.203 &     0.588 & 1.36\\[1.5mm]
  546G09   &    2.25 &      4.00 &       0.189 &     0.483 & 8.12\\[1.5mm]
  158G15A  &    2.13 &      6.16 &       0.124 &     0.305 & 104 \\
  158G15B  &    3.92 &      3.67 &       0.062 &     0.506 & 32.4\\[1.5mm]
  359G31A  &    2.79 &      3.38 &       0.252 &     0.646 & 1.55\\
  359G31B  &    3.27 &      2.67 &       0.323 &     0.655 & 8.35\\[1.5mm]
  576G59A  &    2.35 &      4.33 &       0.165 &     0.361 & 13.2\\
  576G59B  &    1.76 &      5.04 &       0.105 &     0.396 & 11.9\\[1.5mm]
  114G07A  &    1.19 &      6.79 &       0.086 &     0.234 & 107 \\
  114G07C  &    1.29 &      6.82 &       0.070 &     0.229 & 43.6\\[1.5mm]
  405G06A  &    3.45 &      2.90 &       0.244 &     0.730 & 6.88\\
  405G06B  &    4.00 &      2.40 &       0.265 &     0.647 & 1.43\\
  405G06C  &    1.85 &      6.21 &       0.093 &     0.238 & 5.36\\[1.5mm]
  504G10A  &    3.10 &      3.02 &       0.291 &     0.721 & 1.32\\
  504G10B  &    2.25 &      4.92 &       0.174 &     0.402 & 2.83\\
  \hline
  \end{tabular}
  \end{center}
  \end{table}

\section{Abundance derivations}
Abundances are derived either 'directly' by estimating the electron temperature from oxygen emission lines and then computing abundance ratios from the
relative strength of the corresponding emission lines together with correction factors for the degree of ionisation, or by using calibrated relations between
strong-line fluxes and atomic abundances. When describing the methods below, and in further discussions in this paper, we will be using the following notations 
for the strong-line indices,
\begin{eqnarray}\nonumber
R   & = & [{\rm O}~\textsc{iii}] \lambda 4363
      =  I_{[\rm O~\textsc{iii}] \lambda 4363} /I_{{\rm H}\beta },\\\nonumber
R_2 & = & [{\rm O}~\textsc{ii}] \lambda 3727+ \lambda 3729
      =  I_{\rm [O~\textsc{ii}] \lambda 3727+ \lambda 3729} /I_{{\rm H}\beta },\\
N_2 & = & [{\rm N}~\textsc{ii}] \lambda 6548+ \lambda 6584
      =   I_{\rm [N~\textsc{ii}] \lambda 6548+ \lambda 6584} /I_{{\rm H}\beta },\\\nonumber
S_2 & = & [{\rm S}~\textsc{ii}] \lambda 6717+ \lambda 6731
      =   I_{\rm [S~\textsc{ii}] \lambda 6717 + \lambda 6731} /I_{{\rm H}\beta },\\\nonumber
R_3 & = & [{\rm O}~\textsc{iii}] \lambda 4959+ \lambda 5007
      =   I_{{\rm [O~\textsc{iii}]} \lambda 4959 + \lambda 5007} /I_{{\rm H}\beta }.
\end{eqnarray}

We derive abundances for 23 H~\textsc{ii}-regions in LSBGs from the \citet{Ronnback95} LSBG sample where sulfur and nitrogen lines were detected. The
$N_2$ and $S_2$ indices {(see Table \ref{indices})} can then be used to estimate abundances and electron temperatures. We use the 'NS-calibration' by 
\citet{Pilyugin11} and the 'ON-calibration' by \citet{Pilyugin10}. More precisely, we derive abundances using relations of the form
\begin{equation}
12+\log {\rm \left({X\over H}\right)}_{\rm NS}  = a_0 +  a_1\,\log R_3 + a_2\,\log N_2 + a_3\,\log \left({N_2\over S_2}\right),
\label{equation:z}
\end{equation}
and
\begin{equation}
12+\log {\rm \left({X\over H}\right)}_{\rm ON}  = a_0 +  a_1\,\log R_3 + a_2\,\log R_2 + a_3\,\log \left({N_2\over R_2}\right),
\label{equation:z}
\end{equation}
respectively, where the coefficients $a_i$ are fitting parameters obtained using a "calibration sample" of local galaxies with well-measured $T_{\rm e}$-based 
abundances in their H~\textsc{ii}-regions \citep[see][for further details]{Pilyugin10,Pilyugin11}. 
Both the NS- and the ON-calibrations avoid the often weak ${\rm [O~\textsc{III}]} \lambda 4363$-line, and in the case of the NS-calibration
also the ${\rm [O~\textsc{II}] \lambda 3727+ \lambda 3729}$-lines, which are sometimes uncertain in LSBGs (combining all relevant uncertainties one finds that the
$R_2$-index may have a $\pm 20$\% uncertainty). Whenever possible, we also derive 'direct' $T_{\rm e}$-based abundances according to \citet{Pilyugin10}.

The NS-calibration uses the $N_2$ and $S_2$ indices as a kind of 'substitute' for the electron temperature and the ionisation factor, while the ON-calibration 
uses the $N_2$ and $R_2$ indices. However, the $R_2$ index is sometimes hard to determine in LSBGs. Two such examples are ESO 489-G56 and ESO 546-G34 
\citet[see Fig. 2 in][]{Ronnback95} which are both most likely a very metal poor LSB galaxies. Using a calibration without $R_2$, such as the NS-calibration, is 
therefore important to get a good handle on the abundances in some objects. This is especially true for the N/O ratio.

For a smaller number (13) of {H~\textsc{ii}-regions} where the ${\rm [O~\textsc{III}]} \lambda 4363$-line can be detected, we can also derive electron temperatures
$T_{\rm e}$ and corresponding $T_{\rm e}$-based abundances. More precisely, {the O$^+$- and O$^{++}$-zone temperatures, $t_3$ and $t_2$,} are obtained 
iteratively using recent calibrations by \citet{Pilyugin10} based on the principles described in
\citet{Pilyugin09}, i.e.,
\begin{equation}
t_3 = {1.467 \over \log\left({R_3/R}\right)-0.876-0.193\log(t_3)+0.033\,t_3}
\end{equation}
and 
\begin{equation}
t_2 = 0.314+0.672\,t_3, 
\end{equation}
where $t_3$ and $t_2$ are given in units of 10$^4$K. The oxygen and nitrogen abundances can then be calculated from
\begin{equation}
      \log\left({{\rm O}\over {\rm H}}\right) = \log({\rm O}_3+{\rm O}_2),
\end{equation}
where
\begin{equation}
   \log({\rm O}_2) = \log(R_2)+5.929+{1.617\over t_2}-0.568\log(t_2)-0.008\,t_2
\end{equation}
\begin{equation}
      \log({\rm O}_3)=\log(R_3)+6.251+{1.204\over t_3}-0.613\log(t_3)-0.015\,t_3,
\end{equation}
and
\begin{equation}
      \log\left({\rm {N}\over \rm{O}}\right) = \log\left({N_2\over R_2}\right)+0.334-{0.724\over t_2}-0.035 \log(t_2)+0.005\,t_2.
\end{equation}

The emission line data taken from \citet{Ronnback95} has been corrected for extinction and, more importantly, absorption features due to the underlying stellar
population. The underlying absorption in the Balmer lines is tightly correlated with the strength of the 4000\AA -break \citep{Mattsson09}. In terms of the
$D(4000)$-parameter \citep{Bruzual83} the underlying absorption has a maximum just above $D(4000) = 1$ (no or weak discontinuity). We find that all { 23 
H~\textsc{ii}-regions} 
considered in the present study have $D(4000)$ close to unity and thus significant underlying absorption in H$\alpha$ and H$\beta$ is expected. Quantitatively,
this means the correction is around 3\AA\ in the equivalent width of H$\alpha$ and 5\AA\ in H$\beta$. We chose to keep the original corrections {(listed in Table 
\ref{abundances})} made by \citet{Ronnback95} since reanalysis of a few objects (ESO 146-G14, ESO 158-G15 and ESO 546-G34) gave essentially the same 
result and the grid of SEMs used by \citet{Ronnback95} covers a larger part of parameter space\footnote{We used a more up to date grid of spectral evolution 
models (SEMs) computed using the code by \citet{Bruzual03} and applied the same iterative scheme for correction as in \citet{Ronnback95}, but now using 
$D(4000)$ as an additional parameter. This correction scheme is a development of the preliminary results from \citet{Mattsson09} and will be discussed in detail 
in a forthcoming paper. A new larger grid of models is also under development.}. {The software used for applying the original corrections was in itself
corrected after \citet{Ronnback95} had published their results. Hence, the numbers presented here differ slightly from those in \citet{Ronnback95} in some cases.}

  \begin{table*}
  \begin{center}
  \caption{\label{abundances} Derived abundances, {using the notations ${\rm O/H} = 12 + \log({\rm O/H})$ and ${\rm N/O} = \log({\rm N/O})$ for brevity, and 
  corrections for underlying absorption ($W_{\rm abs}$) in \AA ngstr\"oms} for the 23 H~\textsc{ii}-regions. All numbers are based on data from \citet{Ronnback95}, 
  except for ESO 146-G14 where data are taken from \citet{Bergvall95}. {$R_{23}$-abundances are according to \citet{Ronnback95} using the calibration by 
  \citet{McGaugh94}.}}
  \begin{tabular}{lccccccccccc}
  Object  
  & $({\rm O/H})_{\rm NS}$ 
  & $({\rm O/H})_{\rm ON}$
  & $({\rm O/H})_{T_{\rm e}}$
  & $({\rm O/H})_{T_{\rm e}}^{\rm RB95}$
  & $({\rm O/H})_{R_{23}}$
  & $({\rm N/O})_{\rm NS}$    
  & $({\rm N/O})_{\rm ON}$    
  & $({\rm N/O})_{T_{\rm e}}$    
  & $({\rm N/O})_{T_{\rm e}}{\rm RB95}$ 
  & $W_{\rm abs}({\rm H}\alpha)$
  & $W_{\rm abs}({\rm H}\beta)$\\
  \hline
%  146G14A &     7.96 &  7.96 &  -    &  -    &  7.92 & -1.45 &  -1.32 &  -      &  -     & 3.9 & 4.6\\
  146G14B &     7.83 &  7.83 &  7.69 &  7.59 &  8.02 & -1.48 &  -1.42 &  -1.38  &  -1.51 & 5.0 & 6.5\\
  146G14C &     7.93 &  7.91 &  7.60 &  7.54 &  -    & -1.48 &  -1.46 &  -      &  -     & 4.2 & 4.8\\[1.5mm]
  489G56A &     7.59 &  7.60 &  7.47 &  7.49 &  7.63 & -1.50 &  -1.41 &  -1.37  &  -1.35 & 4.0 & 4.8\\
  489G56B &     7.49 &  7.52 &  -    &  -    &  7.52 & -1.46 &  -1.30 &  -      &  -     & 4.2 & 6.3\\[1.5mm]
  505G04A &     7.77 &  7.75 &  7.68 &  7.63 &  8.03 & -1.50 &  -1.52 &  -1.48  &  -1.44 & 4.7 & 5.8\\
  505G04B &     7.82 &  7.81 &  7.82 &  7.77 &  8.01 & -1.47 &  -1.47 &  -1.47  &  -1.43 & 4.5 & 5.6\\[1.5mm]
  546G34A &     7.40 &  7.38 &  -    &  -    &  7.69 & -1.55 &  -1.60 &  -      &  -     & 4.9 & 6.6\\
  546G34B &     7.26 &  7.25 &  -    &  -    &  7.64 & -1.52 &  -1.56 &  -      &  -     & 3.3 & 4.6\\[1.5mm]
  462G22  &     7.82 &  7.82 &  -    &  -    &  7.99 & -1.47 &  -1.41 &  -      &  -     & 4.5 & 6.6\\[1.5mm]
  546G09  &     7.96 &  7.96 &  -    &  -    &  7.97 & -1.46 &  -1.35 &  -      &  -     & 2.6 & 2.7\\[1.5mm]
  158G15A &     8.02 &  8.00 &  8.10 &  8.07 &  8.14 & -1.46 &  -1.47 &  -1.53  &  -1.49 & 1.8 & 3.8\\
  158G15B &     7.60 &  7.53 &  -    &  -    &  8.24 & -1.64 &  -1.89 &  -      &  -     & 3.7 & 4.5\\[1.5mm]
  359G31A &     7.97 &  7.98 &  -    &  -    &  8.05 & -1.45 &  -1.33 &  -      &  -     & 3.2 & 4.0\\
  359G31B &     7.95 &  7.95 &  -    &  -    &  8.05 & -1.41 &  -1.31 &  -      &  -     & 3.8 & 4.7\\[1.5mm]
  576G59A &     7.96 &  7.95 &  8.14 &  8.09 &  8.02 & -1.44 &  -1.41 &  -1.49  &  -1.46 & 3.1 & 3.3\\
  576G59B &     7.88 &  7.88 &  7.77 &  7.81 &  8.00 & -1.52 &  -1.45 &  -1.43  &  -1.47 & 2.9 & 3.0\\[1.5mm]
  114G07A &     7.95 &  7.95 &  8.01 &  8.04 &  8.00 & -1.48 &  -1.38 &  -1.42  &  -1.41 & 2.7 & 3.0\\
  114G07C &     7.89 &  7.89 &  8.05 &  8.03 &  8.04 & -1.51 &  -1.47 &  -1.55  &  -1.52 & 2.8 & 3.2\\[1.5mm]
  405G06A &     7.90 &  7.89 &  7.89 &  7.83 &  8.10 & -1.47 &  -1.42 &  -1.40  &  -1.36 & 4.1 & 5.1\\
  405G06B &     7.85 &  7.83 &  -    &  -    &  8.16 & -1.44 &  -1.44 &  -      &  -     & 2.9 & 3.7\\
  405G06C &     7.94 &  7.92 &  8.14 &  8.11 &  8.09 & -1.47 &  -1.51 &  -1.61  &  -1.59 & 2.6 & 2.8\\[1.5mm]
  504G10A &     7.97 &  7.97 &  -    &  -    &  8.12 & -1.45 &  -1.32 &  -      &  -     & 3.9 & 5.5\\
  504G10B &     8.02 &  8.02 &  8.14 &  8.11 &  8.05 & -1.45 &  -1.38 &  -1.44  &  -1.44 & 5.7 & 7.3\\
  \hline
  \end{tabular}
  \end{center}
  \end{table*}

\section{Results and discussion}
\subsection{Abundances}
The full set of derived abundances, together with the original numbers derived by \citet{Ronnback95}, is presented in Table \ref{abundances}. We estimate the
typical error (mainly due to the method itself) in abundances derived using the NS- and ON-calibrations to be $\pm 0.1$ dex, and the $T_{\rm e}$-based abundances is 
expected to have slightly smaller total errors. {The typical relative mean errors of the line intensities are 5-10\%, which makes the strong-line calibration itself the
dominant source of uncertainty. However, we would like to point out a caveat to the [N~\textsc{ii}]-fluxes in general: the resolution of the \citet{Ronnback95} spectra are
not high enough to resolve the [N~\textsc{ii}]- and H$\alpha$-line in a satisfactory way. This may add further uncertainty to the line fluxes (see the detailed discussion
about ESO 546-G34 below) and typically lead to overestimated $N_2$-indices.}

\begin{figure}
  \resizebox{\hsize}{!}{
   \includegraphics{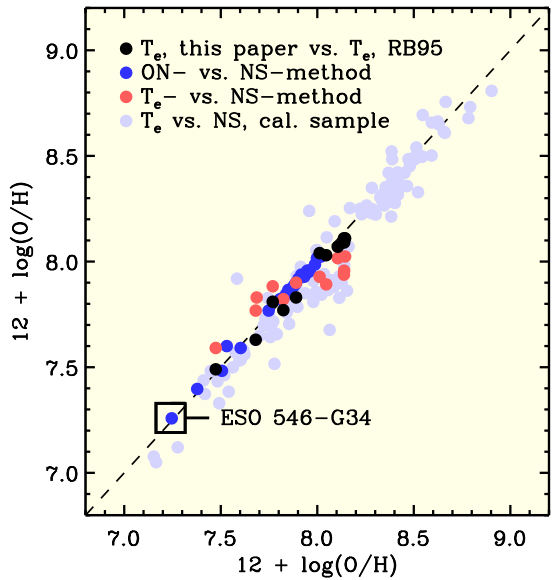}
  }
  \caption{\label{OHcomp} Comparison of oxygen abundances derived using different methods and calibrations. The agreement between the abundances is typically 
  within 0.1 dex with only a few exeptions.}
  \end{figure}

\begin{figure}
  \resizebox{\hsize}{!}{
   \includegraphics{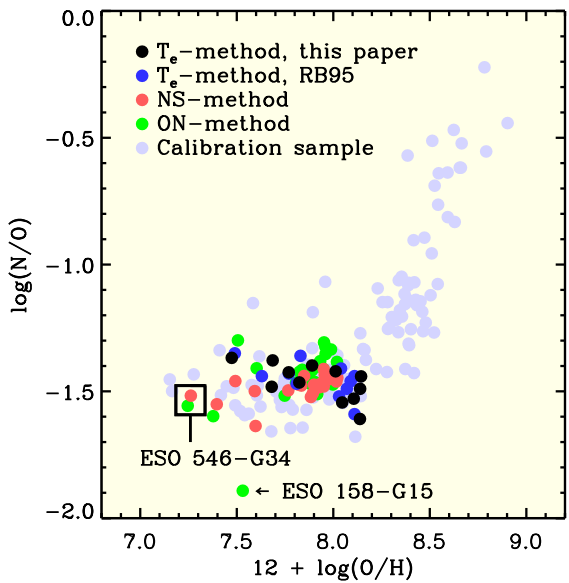}
  }
  \caption{\label{OHNO} N/O-O/H relation as derived using different methods and calibrations. The deviating object is H~\textsc{ii} region B in ESO 158-G15.}
  \end{figure}

The $T_{\rm e}$-based abundances agree very well with the abundances derived by 
\citet[][see also Figs. \ref{OHcomp} and \ref{OHNO} in the present paper]{Ronnback95}. Among the galaxies with detectable ${\rm [O~{\textsc III}]} \lambda 4363$-lines, the
lowest $T_{\rm e}$-based oxygen abundance is found in ESO 489-G56 where $12+\log({\rm O/H}) = 7.47$ \citep[7.49 according to][]{Ronnback95}. In fact, we have only
been able to find one LSBG with lower $T_{\rm e}$-based oxygen abundance when searching the literature 
\citep[CGCG 269-049, $\log({\rm O/H})+12=7.43$,][]{Kniazev03}. 

The ON-abundances also agree with the $T_{\rm e}$-based abundances as well as with the NS-abundances and the existence of a N/O-plateu at low metallicity (see Fig.
\ref{OHNO}). There is essentially only one exception: H~\textsc{ii} region B in ESO 158-G15 which has a significantly lower N/O-ratio than the rest of the sample 
{when using the ON-calibration}. This may {partly} be due to the fact that the $N_2$-value for this object lies on the boundary between the cool and warm regimes 
in the NS- and ON-calibrations, {but also that the $N_2/R_2$-ratio is clearly lower than for the rest of the sample (see Table \ref{indices}).}
Among the oxygen abundances derived using the NS- and ON-calibrations we find one particular H~\textsc{ii} region with a very low abundance. 
This is discussed in detail below.

\subsection{ESO 546-G34}
Among the galaxies in this study, ESO 546-G34 stands out as the most metal poor according to our analysis {($\log({\rm O/H})_{\rm NS}+12=7.26, 7.40$ and
$\log({\rm O/H})_{\rm ON}+12=7.25, 7.38$ for the two H~\textsc{ii}-regions, respectively). This galaxy is a local (redshift $z = 0.00523$) dwarf galaxy of Magellanic type, which 
has an absolute $B$-band magnitude of $M_B = -15.7$, a central surface brightness $\mu_{0, \rm B} = 22.9$ and $B-V = 0.38$ \citep{Ronnback94}. The H~\textsc{i} mass is estimated to 
be $5.4\cdot 10^8 M_\odot$ and the total baryon mass is roughly $1.1\cdot 10^9 M_\odot$ (previously unpublished data, to be presented in more detail in a forthcoming  publication).} 
Previous empirical estimates of the oxygen abundances in two of the three observed H~\textsc{ii}-regions were significantly higher \citep{Ronnback95}, which was likely due 
to use of ill-determined strong-line calibrations (not suitable for very low metallicities).
However, it was noted that the \citet{Skillman89} calibration derived for metal-poor systems gave $\log({\rm O/H})+12=7.26$ for one of the two detected H~\textsc{ii} regions
(Bergvall \& R\"onnback, unpublished), which is essentially the same value we derive using both the NS- and the ON-calibration (see above and Table \ref{abundances}). 
This places ESO 546-G34 among the very most metal poor galaxies known. ESO 546-G34 may thus be the most metal-poor LSBG ever observed. 

ESO 546-G34 show strong oxygen lines relative to other emission lines, which may be a sign of significant shock heating of the interstellar gas.
\citet{Ronnback95} estimated that the shock contribution to H$\beta$ in ESO 546-G34 is very significant - about $20\%$.  But they also concluded that the corrections 
were only minor in most cases and therefore did not include any corrections for shock contributions in their abundance derivations. This does not seem to be appropriate
for ESO 546-G34, however. Using a set of shock models \citep[see][and references therein]{Ronnback95} we confirm that the best fit to the spectra is obtained for a shock 
contribution to H$\beta$ of $\sim 20\%$. Consequently, we find very strong shock contributions to other lines, e.g., $R_3$ has a 70-75\% contribution from shocks.
This is indeed more than expected, but our preliminary analysis shows that excluding shocks leads to a much worse fit. Taken at face value, the corrections imply that 
the oxygen abundance in both H~\textsc{ii}-regions is $\log({\rm O/H})+12< 7.0$, i.e., less than the extreme values of IZw 18 \citep[see][and references therein]{Kunth00,Ostlin00}, 
DDO 68 and SBS 0335-052W \citep{Izotov07}. On the other hand, shock contributions to the line fluxes may increase the inferred electron temperature, which in principle could
lead to higher abundances (after correction) when using the direct method. Hence, shock corrections can lead to both a lower or a higher abundance depending on the 
method/calibration used for deriving it. It is therefore unclear what the effect of shock contributions is in the present case, although it seems most likely the abundances will be 
corrected downwards. 

In the spectra obtained by \citet{Ronnback95} the $[{\rm N~\textsc{ii}}]$- and H$\alpha$-lines cannot be resolved as individual lines. Hence, there is a potential risk that the 
$N_2$-index is underestimated. We have therefore measured the $[{\rm N~\textsc{ii}}]$- and $[{\rm S~\textsc{ii}}]$-lines in the high resolution VLT/FORS2 spectra obtained by 
\citet{Zackrisson06}, covering the spectral region around H$\alpha$ (5750-7310\AA ). It appears the $[{\rm N~\textsc{ii}}]$-line flux relative to H$\beta$ may be slightly 
underestimated (the H$\alpha$-flux would hence be overestimated) in the study by \citet{Ronnback95}, but the data are still consistent within the uncertainties. For the $[{\rm S~\textsc{ii}}]$-lines 
there is good agreement, however. The $N_2$-index is about 30\% higher in the VLT observations compared to the same in the old data. If this is all due to line blending, then 
$\log({\rm O/H})_{\rm NS}+12=7.34$ and $\log({\rm O/H})_{\rm ON}+12=7.33$ for region B, and $\log({\rm O/H})_{\rm NS}+12=7.47$ and 
$\log({\rm O/H})_{\rm ON}+12=7.46$ for region A (not including any effects of shock heating). We cannot rule 
out the possibility that this may suggest slightly higher abundances, but to be certain new high-resolution spectra (over the whole relevant wavelength range) are 
needed. Nonetheless, the abundances derived here for ESO 546-G34 are under all circumstances unusually low and it is very unlikely that better data will make these 
numbers increase significantly.

  \begin{figure}
  \resizebox{\hsize}{!}{
   \includegraphics{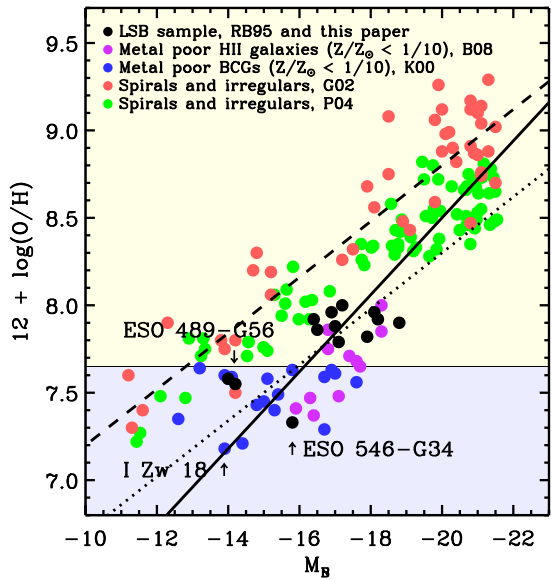}
  }
  \caption{\label{OHMB} Luminosity-metallicity relation for various types of galaxies. LSBGs and BCGs seem to fall outside ($\sim 0.5$ dex below in oxygen
  abundance) the relation for spiral and irregular
  galaxies. The filled black circles show the LSBG sample with {mean} oxygen abundances obtained using the NS-calibration and B-band magnitudes taken from RB95,
  {while the purple and blue filled circles show metal-poor BCGs and H~\textsc{ii} galaxies taken from the literature.
  The full line is a fit to the LSBGs and spiral galaxies (green circles with $M_B$ brighter than $-17$) from \citet{Pilyugin04}. The dashed line shows the fit to the
  spiral and irregular galaxies  derived by \citet[][red circles]{Garnett02}; the dotted line show the same relation scaled down by 0.5 dex. Objects falling on the
  blue-shaded area have metallicities below 1/10 of solar ($12+\log({\rm O/H}) \le 7.65$), usually referred to as the extremely metal-poor regime.}}
  \end{figure}

\subsection{Luminosity-metallicity relation and the BCG-LSBG connection}
BCGs (or H~\textsc{ii} galaxies in general) may be triggered by mergers involving gas-rich galaxies with little star formation, e.g., LSBGs 
\cite{Bergvall99,Bergvall00,Taylor96,Taylor97}. In Fig. \ref{OHMB} we show the {oxygen abundances} we derive for the LSBGs {using the NS-method} in 
a O/H vs. absolute B-magnitude diagram. Very metal-poor galaxies and LSBGs appear to fall below the usual 
luminosity-metallicity trend for spiral and irregular galaxies. For similar magnitudes $\log({\rm O/H})$ is 
typically 0.5 dex lower in our LSBG sample as well as in very metal-poor BCGs \citep[data taken from][]{Kunth00} and the H~\textsc{ii} sample by \citet{Brown08}. 
This may lend support to the idea of an evolutionary 
connection between LSBGs and BCGs, although the data is also consistent with a single relation with large scatter.

In case LSBGs and BCGs fall outside the regular luminosity-metallicity relation, and there is an evolutionary connection between LSBGs and BCGs, {what 
would this connection be? A most likely scenario is the one} where BCGs are the result of a major merger between a blue LSB galaxy and another (but evolved) 
galaxy of similar size, possibly a dwarf elliptical, thus triggering a starburst \citep{Bergvall99}.
This hypotheses naturally explains the LSB component detected in many BCGs \cite[see, e.g.][]{Bergvall02} and the similarities in the colours between BCGs
and blue LSBGs \citep{Telles97}. Moreover, it also have an interesting implication: if there 
are very metal-poor BCGs, there should be equally (or more) metal-poor LSBGs to be found.

\section{Conclusions}
We have reanalysed spectroscopic data for 23 H~\textsc{ii}-regions in blue, metal-poor LSBGs taking advantage of recent advances in calibrating strong-line
methods by \citet{Pilyugin10} and \citet{Pilyugin11}. We have identified a galaxy (ESO 546-G34) which may be the most metal-poor LSB galaxy in the local Universe.
We are reasonably sure this is an extremely metal-poor galaxy, although it would be necessary to confirm the abundances using 'direct' methods, i.e., 
$T_{\rm e}$-based abundance determinations. For this to be possible ESO 546-G34 needs to be re-observed with better signal-to-noise, which is something we hope to
be able to return to. 

Finally, we think there is now evidence that blue metal-poor LSBGs, together with many BCGs and H~\textsc{ii} galaxies, fall outside the regular luminosty-metallicity 
relation. This suggests there might be an evolutionary connection between LSBGs and BCGs to be found, as suggested by \citet{Bergvall99,Bergvall00}. In such a case,
there should be several very metal-poor LSBGs in the local Universe. We would hence like to encourage the community to search for such candidates.

\section*{Acknowledgments}
The authors wish to thank the anonymous reviewer for his/her constructive criticism that helped to improve this paper.
L.S.P. acknowledges support from the Cosmomicrophysics project of
the National Academy of Sciences of Ukraine. L.M. acknowledges support from
the Swedish Research Council. The Dark Cosmology Centre is funded by the Danish National Research Foundation.

\end{document}